\documentclass[%
 reprint,
 amsmath,amssymb,
 aps,
]{revtex4-1}

\usepackage{graphicx}
\usepackage{dcolumn}
\usepackage{bm}
\usepackage{graphics}
\usepackage{bm}
\usepackage{amsmath}
\usepackage{epstopdf}
\usepackage{amsmath}
\usepackage{amsfonts}
\usepackage{color}
\usepackage[]{algorithm2e}
 \usepackage{tikz}
 \usepackage{mathtools}
 \usepackage[percent]{overpic}
 \usepackage{float}
 \usepackage{subfig}
 \usepackage{comment}
\let\MYoriglatexcaption\caption
\renewcommand{\caption}[2][\relax]{\MYoriglatexcaption[#2]{#2}}


\renewcommand{\eqref}[1]{(\ref{#1})}

\begin{document}

\preprint{APS/123-QED}

\title{Random Walk with Memory on Complex Networks} 

\author{Lasko Basnarkov$^{1,2}$}
\email{lasko.basnarkov@finki.ukim.mk}
\author{Miroslav Mirchev$^{1}$}
\author{Ljupco Kocarev$^{1,2}$}

\affiliation{$^{1}$Faculty of Computer Science and Engineering, SS. Cyril and Methodius University, P.O. Box 393, 1000 Skopje, Macedonia\\}%

\affiliation{$^{2}$Macedonian Academy of Sciences and Arts, P.O. Box 428, 1000 Skopje, Macedonia}%


\date{\today}

\begin{abstract}

We study random walk on complex networks with transition probabilities which depend on the current and previously visited nodes. By using an absorbing Markov chain we derive an exact expression for the mean first passage time between pairs of nodes, for a random walk with a memory of one step. We have analyzed one particular model of random walk, where the transition probabilities depend on the number of paths to the second neighbors. The numerical experiments on paradigmatic complex networks verify the validity of the theoretical expressions, and also indicate that the flattening of the stationary occupation probability accompanies a nearly optimal random search.

\end{abstract}

\maketitle

\section{Introduction}

The pursuit for appropriate models of the nontrivial interconnections between the units of real systems has led to the emergence of the complex networks theory as one of the most fruitful fields in modern science. Instead of being regular, or purely random \cite{erdos1960evolution}, the graph of connections between the items rather frequently possesses characteristics like the small world property \cite{watts1998collective} and power law degree distribution \cite{barabasi1999emergence}. These topological features have strong implications on the dynamics which might be present in the system. A list of such dynamical processes on complex networks of interacting units can include synchronization \cite{arenas2008synchronization}, consensus formation \cite{olfati2007consensus}, disease spreading \cite{newman2002spread} and so on. 

The random walk is one of the most pervasive concepts in natural sciences which is applied in studies of diverse phenomena ranging from simple animal strategies for food location \cite{hughes1995random, viswanathan1999optimizing} to complex human interactions resulting in stock price variations \cite{bachelier1900theorie}, or evolution of research interests \cite{jia2017quantifying}. A recent paper \cite{masuda2017random} contains nice review of the topic and long list of references. Large portion of dynamical processes on complex networks like the PageRank algorithm \cite{brin1998anatomy}, various types of searching \cite{carmi2006searching,boguna2009navigability}, or community detection \cite{rosvall2008maps} are based on or related to the random walk. Random searching process in a complex network is formulated as follows: starting from an arbitrary node, or source $i$, sequentially hop from a node to one randomly chosen neighbor until reaching some previously defined target node $j$. The performance of a searching procedure is measured in terms of the number of steps needed to get from $i$ to $j$ and the related quantity is known as first passage time. Due to the stochastic nature of picking the nodes in the sequence, sometimes one can be very lucky and rather quickly find the target, while in most of the trials the number of steps would be larger then the number of nodes in the network, for a typical source-target pair. Therefore, a more informative quantity is the average number needed to complete the task -- the Mean First Passage Time (MFPT) -- obtained by averaging across all possible realizations of the random choices. 

On the other side, there are efficient deterministic searching algorithms, which rely on information about the underlying graph structure. In such approaches, when one has knowledge of the full structure of the graph, the shortest paths are used, and then one needs smallest number of steps to reach the target. However, for very large systems, like the World Wide Web, or in dynamical environments like mobile sensor networks, keeping and updating all necessary topological information might be serious issue. Then one could turn towards strategies based on local information only. The classical Uniform Random Walk (URW) needs the smallest amount of information -- only the number of neighbors (the degree $k_i$) of each node $i$. Within this approach, the probabilities for choosing among the neighbors of some node $i$ are taken to be identical and equal to the inverse of its degree $p = 1 / k_i$. However, this procedure greatly increases the time to completion of the task, which is another type of inconvenience. The searching can be improved when the local information extends the node degrees. For example, it was shown that for a certain type of small world networks, random target can be found rather quickly by using local information only \cite{kleinberg1999small, kleinberg2006complex}. Knowledge of the identities of the direct or maybe more distant neighbors, also enhances the searching \cite{borgs2012power}.

There are various alternatives for modification of the URW aimed for speeding up its searching capabilities. Some of these works provided enhancements while others also presented connections with related problems in other fields. For example, as a counterpart of the path integrals, the Maximal Entropy Random Walk was introduced as a modification of URW which assigns equal probabilities to all paths with equal length starting from a certain node \cite{burda2009localization}. In another approach, the L\'{e}vy random walk which allows for jumps toward more distant nodes besides the (first) neighbors, was proven to decrease the expected time needed to visit all nodes in a network \cite{riascos2012long}. Combination of the local diffusion and knowledge of the topology has recently been applied for study of routing of neural signals \cite{avena2019spectrum}. Biasing of the random walk has been shown to be useful in sampling of complex networks as well \cite{shioda2014random}. Another important achievement was the demonstration that biasing of the URW, by preferring the less connected neighbors, can improve the random searching in complex networks \cite{fronczak2009biased}. In the same contribution, it was uncovered that inverse-degree-based biasing of the random walk also leads to uniform stationary occupation probability. In a related work, it was obtained that the improvement is greatest when the probability to jump to a neighbor is inversely proportional to its degree \cite{bonaventura2014characteristic}. 

In this work we explore the potential for searching improvement by considering memory-based random walk on complex networks since it relies on information that extends the immediate neighborhood. First, we develop theoretical framework for analytical calculation of the MFPT between any pair of nodes when the random walk has a memory of one step. Then we apply it for determination of MFPT for one particular searching algorithm which aims to provide nearly equal chances of visiting second neighbors. We numerically show that searching enhancement is also accompanied with flattening of the stationary distribution of the visiting frequency as it is the case of the inverse-degree-based biasing of the random walk. The co-occurrence of the improved searching and nearly uniform stationary distribution is found even for the memory-based and inverse-degree-based random walk on directed complex networks as well.

The remainder of the text is organized as follows. In Section \ref{SEC:with_mem} we present general theoretical framework for studying random search with random walk with memory. Then in Section \ref{sec:algorithm} we introduce an algorithm for random search with memory of one step. The results from the numerical experiments and their analysis are provided in Section \ref{SEC:results}. The paper finishes with the Conclusions.

\section{Mean First Passage Time of Random walk with memory on complex networks }\label{SEC:with_mem}

Consider a connected complex network with $N$ nodes, with adjacency matrix $\mathbf{A}$. We will study discrete-time walk, where the next node in the sequence is chosen randomly, with time-invariant transition, or jump probability which depends on the previously visited nodes. Our theoretical analysis will be focused on the simplest scenario with memory, when this probability depends only on the present and the node visited immediately before it. This situation can be identified as one with a memory of depth (or length) one. To be more specific, assume the random walker at certain time step has moved from node $r$ to its neighbor $s$. For one-step memory, the probability of proceeding towards some neighbor $t$ from $s$ \footnote{One should note that for directed networks, the neighbor $t$ must be chosen among those towards which $s$ points to.}, depends only on the previously visited node $r$, but not on the preceding ones. It is thus given with 
\begin{equation}
    p(t|s, r, u_1, u_2, \dots) = p(t|s, r),
\end{equation}
where $u_1, u_2, \dots$ denotes the sequence of nodes visited before $r$. This kind of random walk can be suitably studied with a related Markov chain with states that represent the pairs of neighboring nodes in the network. To make the connection between the random walk and its associated Markov chain more intuitive, let us denote with $rs$ the state in the Markov chain when the random walker has visited node $r$ immediately before $s$. The transition probabilities in the chain from state $rs$ to $st$ are thus $p_{rs, st} = p(t|s, r)$. All such transition probabilities can be compactly organized in the respective transition probability matrix $\mathbf{P}$. We note that although the generalization of the associating Markov chain for a random walk with longer memory is straightforward, one should keep in mind that the size of the corresponding matrix will rise exponentially. This is due to the fact that the states of such a Markov chain will consist of all allowable sequences of successively visited nodes with length that equals the memory depth plus one.

When the transition probability matrix of the related Markov chain is determined, such a random walk will be completely defined once the starting step is specified. One particular initialization of the walk which starts from some node $i$ is to choose randomly one of its neighbors and then continue with the memory-based algorithm specified with the transition matrix $\textbf{P}$. Finding some target $j$ in the network corresponds to reaching any of the states denoted with $sj$ in the Markov chain, where $s$ is any neighbor of the node $j$ \footnote{For directed network, nodes $s$ are only those which point to $j$}. Then, the MFPT from node $i$ to $j$ could be related to the Mean Time to Absorption (MTA) of a random walk in properly chosen absorbing Markov chain that started in any state $ir$. In that chain, all states $sj$, where $s$ is neighbor of the target $j$, are absorbing, while the remaining ones $rt$, where $t \neq j$, and $r \neq j$ are transient states. We also note, that in the Markov chain that models the random walk with memory there can be states denoted as $js$. They can be included in the absorbing chain only if one needs to calculate the Mean Recurrence Time, or the average time needed for the walker starting from $j$ to return at $j$ again. When the starting node differs from the target, such states can be omitted in order to reduce the size of the matrices involved. Before deriving the relationship between MTA and MFPT, we will first present some well known results about the MTA in absorbing Markov chains, which can be found for example in \cite{grinstead2012introduction}.
For such purpose, one should first determine the transition matrix of the absorbing Markov chain, which depends on the target $j$, and thus we will denote it with $\mathbf{P}_{(j)}$. Since the random walk should stop at any absorbing state, the probability of leaving any of them is zero. Also, the transition probabilities between the other, the transient states in the absorbing chain, are the same as the respective ones in the original chain. Thus, the absorbing chain matrix $\mathbf{P}_{(j)}$ differs from the general matrix $\mathbf{P}$ only in the rows with index $sj$, which in the absorbing matrix have values $p_{sj, rt} = \delta_{sj, rt}$. The transition matrix of the absorbing Markov chains is conveniently represented in the canonical form. For memory-based random walk targeting the node $j$ the canonical form reads
\begin{equation}
\label{eq:P_j}
    \mathbf{P}_{(j)} = \begin{vmatrix}
\mathbf{Q}_{(j)}&\mathbf{R}_{(j)}\\
\mathbf{0}&\mathbf{I}\\
\end{vmatrix}.
\end{equation}
In the last equation $\mathbf{Q}_{(j)}$ is a matrix consisting of transition probabilities between the transient states. The submatrix $\mathbf{R}_{(j)}$ is determined with the rows corresponding to the transient states, while the columns are indexed with all absorbing states $sj$, which are related to the target $j$. We remind that the transition probabilities in these two matrices have identical values to the respective ones in the original chain. Lastly, the appropriately sized zero matrix $\mathbf{0}$ and the identity matrix $\mathbf{I}$ denote that from any of the absorbing states $sj$ the random walker does not pursue further and remains in the same state. To simplify the notation, we will use Greek letters $\alpha$ and $\beta$ to identify states in the absorbing Markov chain, instead of using pairs of neighboring network nodes.

The MTA equals the average number of steps while the walker is in the transient states. The probabilities of presence in the transient states is encoded in the powers of the transient submatrix $\mathbf{Q}_{(j)}^n$. Its terms $q_{(j), \alpha, \beta}^{(n)}$ are the probabilities of the walker which started at state $\alpha$ to be at $\beta$ after $n$ steps. Let us introduce a binary random indicator variable $I_{(j),\beta}^{(n)}$, which has value 1, if the walker is present at state $\beta$ at moment $n$, and 0 if it is absent. Its expected value $E(I_{(j),\beta}^{(n)})$ equals the probability $q_{(j),\alpha,\beta}^{(n)}$. Then, the expected number of steps when the walker starting at $\alpha$ is present at $\beta$ in the first $n$ iterations is
\begin{eqnarray}
    E(I_{(j),\beta}^{(0)} + I_{(j),\beta}^{(1)} + \cdots + I_{(j),\beta}^{(n)}) \nonumber\\= q_{(j), \alpha,\beta}^{(0)} + q_{(j), \alpha,\beta}^{(1)}+ \cdots + q_{(j), \alpha,\beta}^{(n)}.
\end{eqnarray}

The expected number of visits of the state $\beta$, for an infinite walk is obtained by simply letting $n \to \infty$. One can introduce a fundamental matrix for this absorbing Markov chain as the infinite sum 
\begin{equation}
    \mathbf{Y}_{(j)} = \mathbf{I} + \mathbf{Q}_{(j)} +\mathbf{Q}_{(j)}^2 + \cdots.
\end{equation}
It contains the expected number of steps in which perpetual random walk starting from any state $\alpha$ (the row) is present at node $\beta$ (the column). The MTA of random walk which started at state $\alpha$ equals the mean number of steps in which the walker is in {\it any} transient state $\beta$, or the sum
\begin{equation}
    \mu_{(j), \alpha} = \sum_{\beta} y_{(j), \alpha,\beta},
\end{equation}
where $y_{(j), \alpha, \beta}$ are the elements of the fundamental matrix of the absorbing chain $\mathbf{Y}_{(j)}$. A more compact expression can be obtained by using the vector $\mu_{(j)}$ consisting of all MTA $\mu_{(j),\alpha}$ from all possible starting states $\alpha$, by using the matrix equation
\begin{equation}
    \mu_{(j)} = \mathbf{Y}_{(j)}\mathbf{c},
\end{equation}
where $\mathbf{c}$ is a column vector with all elements equal to one. Because any random walk finishes in some absorbing state with probability one, the powers of the matrix $\mathbf{Q}_{(j)}^{(n)}$ become vanishingly small as $n\to\infty$. Accordingly, the infinite sum of matrices converges and the fundamental matrix can be represented more compactly as 
\begin{equation}
    \mathbf{Y}_{(j)} = \left(\mathbf{I} - \mathbf{Q}_{(j)}\right)^{-1}.\label{eq:Fund_mat_absorbing}
\end{equation}
The last relationship provides efficient procedure for determination of the MTA based on calculation of inverse matrix. As we will see below, the MTA vector $\mu_{(j)}$ contains sufficient information for calculation of MFPTs from all starting nodes to particular target $j$. We note that in alternative interpretation the MTA is average number of steps needed for the walk to finish in any absorbing state, and is thus an average over all possible absorbing states.

The random walks on complex networks are such that at a single time step, exactly one hop is made. Then, each random first passage time equals the number of steps needed for reaching the target for the first time, that is the length of the respective random walk. Thus, by definition MFPT between the starting node $i$ and the target $j$ is weighted sum of the lengths $l$ of all walks $\mathcal{W}_{i,j}$, which visit $j$ only at the last step
\begin{equation}
    m_{i,j} = \sum_{\mathcal{W}_{i,j}} l\left(\mathcal{W}_{i,j}\right) p\left(\mathcal{W}_{i,j}\right),
\end{equation}
where $p(\mathcal{W}_{i,j})$ is the probability of occurrence of the walk $\mathcal{W}_{i,j}$. Let us first consider the case when the target is not neighbor of the source. The sum in the last expression can be organized by summing over all walks with memory that visit the neighbor $s$ of $i$ at the first step, and then summing in the whole neighborhood $\mathcal{N}_i$ of $i$ 
\begin{equation}
    m_{i,j} = \sum_{s \in \mathcal{N}_i} p_{i,s} \sum_{\mathcal{W}_{s,j}} \left[1 + l(\mathcal{W}_{s,j})\right] p(\mathcal{W}_{s,j}),
    \label{eq:MFPT_decomposed}
\end{equation}
where $p_{i,s}$ denotes the probability to hop from $i$ to $s$ in the first step. It is known that every random walk in an absorbing Markov chain is absorbed with probability one \cite{grinstead2012introduction}. This implies that the measure of the memory based random walks in the complex network that miss the target $j$ indefinitely, is zero. This practically means that the normalization condition of the probabilities of the memory-based walks that pass through each neighbor $s$ of the initial node $i$ and reach the target $j$, is given as 
\begin{equation}
    \sum_{\mathcal{W}_{s,j}} p(\mathcal{W}_{s,j}) = 1,
    \label{eq:paths_norm}
\end{equation}
where the summation is made for each neighbor $s$ separately. One can also note that the MFTP from the neighbor $s$ of the starting node $i$ to the target $j$ by pursuing the memory-based random walk equals the MTA from the starting state $is$ in the absorbing Markov chain determined with the same target. This MTA is the respective term of the MTA vector $\mu_{(j)}$ and is given with the following sum 
\begin{equation}
    \mu_{(j), is} = \sum_{\mathcal{W}_{s,j}} l(\mathcal{W}_{s,j}) p(\mathcal{W}_{s,j}). \label{eq:MTA_from_is_to_j}
\end{equation}
When the neighbor in the first step is chosen uniformly, one has $p_{i,s} = 1/k_i$. Then, by using (\ref{eq:paths_norm}) and (\ref{eq:MTA_from_is_to_j}) in (\ref{eq:MFPT_decomposed}) one can express the MFPT from the node $i$ to $j$ through the MTAs obtained by the Markov model for the memory-based random walk as
\begin{equation}
    m_{i,j} = 1 + \frac{1}{k_i} \sum_{s \in \mathcal{N}_i} \mu_{(j), is}.
    \label{eq:MFPT_memory_based_final}
\end{equation}

We can now consider the case when the target $j$ is neighbor of $i$. This target could be reached in one step with probability $p_{i,j}=1/k_i$, by pursuing the direct one-step route, while for the mean number of hops through all other walks one can apply the same reasoning as above. We note that in the sum running in the neighborhood of the initial node $i$, the target $j$ should be omitted. Then by using the normalization condition (\ref{eq:paths_norm}), one can obtain that  
\begin{eqnarray}
    \sum_{\substack{s \in \mathcal{N}_i \\ s \neq j}} p_{i,s} \sum_{\mathcal{W}_{s,j}} \left[1 + l(\mathcal{W}_{s,j})\right] p(\mathcal{W}_{s,j}) \nonumber\\ = \frac{k_i - 1}{k_i} + \frac{1}{k_i} \sum_{\substack{s \in \mathcal{N}_i \\ s \neq j}} \mu_{(j), is}.
\end{eqnarray}
Adding the contribution of the direct walk to the last expression, one will obtain similar result as (\ref{eq:MFPT_memory_based_final})
\begin{equation}
    m_{i,j} = 1 + \frac{1}{k_i} \sum_{\substack{s \in \mathcal{N}_i \\ s \neq j}} \mu_{(j), is}.
    \label{eq:MFPT_memory_based_neighbor}
\end{equation}
By using the trivial value $\mu_{(j),ij} = 0$, one can see that the same expression (\ref{eq:MFPT_memory_based_final}) can be used for any target, regardless whether it is neighbor to the starting node, or not.

The analysis in this section can be applied for the simpler case as well -- the random walk without memory. The Markov transition matrix in this situation is consisting of the transition probabilities between the nodes. Then with each target node is associated only one absorbing state -- the target itself. However, it is more convenient to have one fundamental matrix for the whole network, instead of using a different one for each node separately. Without going into details which can be found for example in \cite{grinstead2012introduction}, we will briefly state how it is obtained. For a connected graph with jumping probabilities summarized in transition matrix $\mathbf{P}$, one first determines the row eigenvector $\mathbf{w} = \mathbf{wP}$ that corresponds to the largest eigenvalue. The term $w_j$ of the eigenvector $\mathbf{w}$ represents the probability that the walker will be at node $j$ at infinity. Thus, this vector contains the stationary occupation probabilities or frequency of visits of nodes by a perpetual random walk. Next, one constructs a square matrix $\mathbf{W}$ with identical rows consisting of the vectors $\mathbf{w}$ stacked one on top of another. The respective fundamental matrix for a random walk without memory on a complex network is then given by 
\begin{equation}
    \mathbf{Z} = (\mathbf{I} - \mathbf{P} + \mathbf{W})^{-1},
    \label{eq:Fund_matrix_Z}
\end{equation}
where $\mathbf{I}$ is an identity matrix with the same size as $\mathbf{P}$. The MFPT between the starting node $i$ and the target $j$ is then obtained from the elements of the fundamental matrix, $z_{i,j}$, from the following relationship
\begin{equation}
    m_{i,j} = \frac{z_{j,j}-z_{i,j}}{w_j}.
    \label{eq:MFPT_through_Z}
\end{equation}
The reader interested in a more detailed and intuitive derivation of the same expression (\ref{eq:MFPT_through_Z}) with the generating functions formalism, but for lattices only, can refer to \cite{hughes1995random}, while for complex networks, based on the Laplace transform, deeper explanation can be found in \cite{noh2004random, fronczak2009biased}. We use the last expression for calculation of the MFPT between the nodes in the memoryless random walk with which our model is compared.

One should note that the MFPT is a property of the network parameterized by two nodes -- the starting one $i$ and the final $j$ and is thus sensitive to the choice of this pair. A related property of one node only is obtained by averaging all MFPTs starting from all other nodes and targeting it
\begin{equation}
g_i = \frac{1}{N} \sum_{j=1}^N m_{j,i}.
\label{eq:GMFPT_def}
\end{equation}
In the literature it was called Global Mean First Passage Time -- GMFPT \cite{tejedor2009global}.
This property can be also seen as a kind of centrality measure of nodes in a complex network. By going one step further, one can average across GMFPTs for all nodes and get a property of the whole network which was introduced as Graph MFPT (GrMFPT) \cite{bonaventura2014characteristic}
\begin{equation}
    G = \sum_{i=1}^{N} g_i.
    \label{eq:GrMFPT}
\end{equation}
We use this variable for comparison of the searching by different random walks in complex networks.

\section{Searching algorithm based on random walk with memory of one step} \label{sec:algorithm}

The results for the MFPT obtained in the previous section are general and hold for every random walk with jumping probabilities depending on the current and the previously visited node. They are given in a form that does not provide much intuition about which navigation rules provide better search of the target. Even from the expression for the MFPT of the memoryless walk, one is not sure how the transition probabilities should be defined in order to obtain faster search. We stress that, an interesting contribution was the finding that if the probability to jump to a neighbor is inverse of that neighbor's degree, then the search in undirected network is faster as compared to the URW, and in that scenario the stationary occupation probability approaches the uniform one $\mathbf{w}_j \approx 1/N$  \cite{fronczak2009biased}. This suggests that searching improvement could be expected from biasing which increases the probability for visiting poorly connected nodes, as the inverse degree algorithm does. As shown in the Appendix \ref{appendix}, under certain circumstances inverse-degree biasing can result in nearly constant distribution of visiting frequencies even for memoryless random walk on directed networks as well. This flattening of the invariant density happens in well connected networks, in which each node has many neighbors. As we will see below, our numerical simulations indicated that inverse-degree biasing does not bring searching improvement for networks with small average degree. In that case the distribution of visiting frequency deviates more significantly from the uniform one as well. Thus, navigation rules which favor jumps towards less connected nodes and result in nearly uniform distribution of visiting frequency could be a candidate of a good searching algorithm.

Memory-based algorithms are obviously more complex than memoryless counterparts and their implementation could be justified if they provide improved searching. Guided by the reasoning above, one can pursue a strategy which should result in decreased differences between the probabilities for reaching the second neighbors, which hopefully would bring uniform stationary occupation probability and faster searching. One intuitive way to make such navigation rules is as follows. Assume that at the previous step the walker was at node $r$, from where it has jumped to the node $s$, and in the next step it would visit some node $t$ from the set of neighbors of $s$. Denote the number of all two-hop walks from node $r$ to $t$ with $b_{rt}$. The matrix $\mathbf{B}$ with elements $b_{rt}$ is the square of the adjacency matrix $\mathbf{A}$, $\mathbf{B} = \mathbf{A}^2$. Then, the probability to visit node $t$ after being at nodes $r$ and $s$ in the previous two steps, corresponds to the transition probability from state $rs$ to $st$ in the related Markov chain. In analogy to the inverse-degree biasing, one could favor visiting the less accessible second neighbors by choosing the following jumping probability
\begin{equation}
    p_{rs,st} = \frac{\frac{1}{b_{rt}}}{\sum_{u\in \mathcal{N}_s}\frac{1}{b_{ru}}},
 \label{eq:two_hop_prob_def}
\end{equation}
where the sum in the denominator is used for normalization of the probabilities and it runs in the neighborhood of the node $s$, $\mathcal{N}_s$. This formula assigns a larger weight to nodes $t$ which have less alternative paths to be reached from node $r$, i.e. those with smaller $b_{rt}$. In this way, the probability to visit a node of that kind from $r$ in two steps will be increased, and become closer to that of nodes which are accessible from $r$ in two steps through more alternative ways. We note that for undirected networks every node is a second neighbor to itself, and there is a chance to return to the same node $r$. However, $b_{rr} = k_r$ and the probability $p_{rs,sr}$ is the lowest within all $p_{rs,st}$, hence, the immediate returning is disfavored. In this way, the appearance of short loops is suppressed.

The related Markov model of a random walk with memory could be successfully applied for analytical calculation of the stationary occupation probability as well, which could be used to check whether its flattening  is accompanied by a searching improvement. To find the stationary occupation probability, one should first calculate the invariant distribution of the states of the related Markov chain $\mathbf{v}$, which is obtained from the stationarity condition $\mathbf{vP} = \mathbf{v}$, of the full transition matrix $\mathbf{P}$ of the Markov chain. Its terms are the stationary probabilities of states $v_{rs}$ that correspond to all pairs of neighbors $rs$. Then, the stationary distribution of frequency of visits of the node $s$, by a random walk with memory of one step, would be either of the sums $\sum_r v_{rs}$, or $\sum_t v_{st}$, running within the neighborhood of the node $s$.

\section{Numerical results} \label{SEC:results}

In this section we provide the results obtained by using analytical expressions and numerical simulations with memory-based random walk and compare them with the uniform and inverse-degree-biased random walk. The search effectiveness was studied by calculations of the GrMFPT of each considered network. The stationary occupation probability was also calculated to check whether its flattening accompanies efficient searching. First, we conduct a thorough analysis using generic network models, such as random, scale-free and small world networks. Then, we apply the approaches on two real networks: the Internet at autonomous systems level (undirected), and a reduced set of Wikipedia pages (directed).

The calculations of theoretical expressions involve inverse matrix operation, and the latter presents the major constraint in our analysis. For the random walk with memory, the number of states in the related Markov chain equals the number of links, which limits the size of networks that we could study. Therefore, we have opted to perform the analyses of the MFPT and the invariant density for networks with $N=100$ nodes. We have varied the average node degree, by changing the native model parameters, to see how the connectivity affects the search. For both the analytical and the numerical results, we averaged over $10$ network instances for every parameter setting for each network type. Moreover, in the numerical simulations we have performed $100$ repetitions of the search among all node pairs, for each scenario.

We studied purely random graphs, scale-free and small-world networks as the most typical kinds of networks. For generating such graphs we used algorithms from the NetworkX library in Python which allow construction of the three graph types with given parameter values \cite{NetworkX}. The random graphs are complex networks created according to the Erd\H{o}s-R\'enyi model where every pair of nodes $i$ and $j$ is connected with some predefined probability $p$, which appears as parameter of the graph together with the number of nodes $N$ \cite{erdos1960evolution}. If the probability $p$ is large enough then the obtained graph would very likely be connected -- there will be a path between each pair of nodes. The small world networks were built following the Watts-Strogatz model \cite{watts1998collective}. It starts with a regular ring lattice network with $N$ nodes each connected with $n$ neighbors, and then randomly rewires the links with some probability $p$. The scale free networks were generated using the Barab\'{a}si-Albert model which sequentially builds the network by adding nodes one by one \cite{barabasi1999emergence}. The network builds upon a seed of $m_0$ nodes without edges, and every newly added node forms $m$ links with the existing network \footnote{The parameters $m_0$ and $m$ here are denoted as the authors Barab\'{a}si and Albert originally did and are different from the elements of the MFPT matrix $m_{ij}$.}. Preferential attachment is employed as the probability to connect to an existing node is taken to be proportional to its degree. 

\begin{figure}[!tb]
    \subfloat[BA networks \label{fig:GMFPTBANet}]{
        \includegraphics[width=0.9\columnwidth]{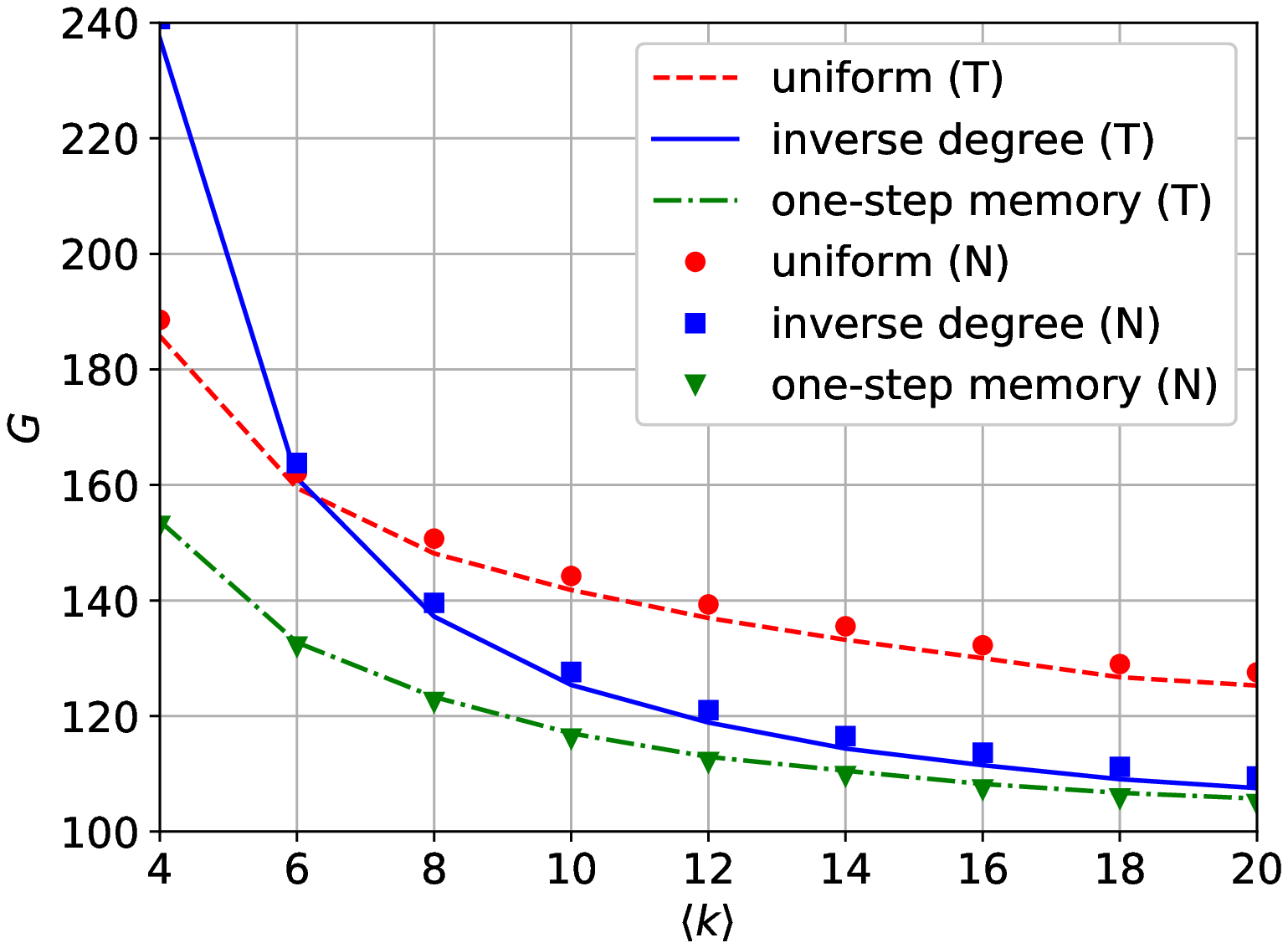}
	}\\
    \subfloat[ER networks \label{fig:GMFPTERNet}]{
    	\includegraphics[width=0.9\columnwidth]{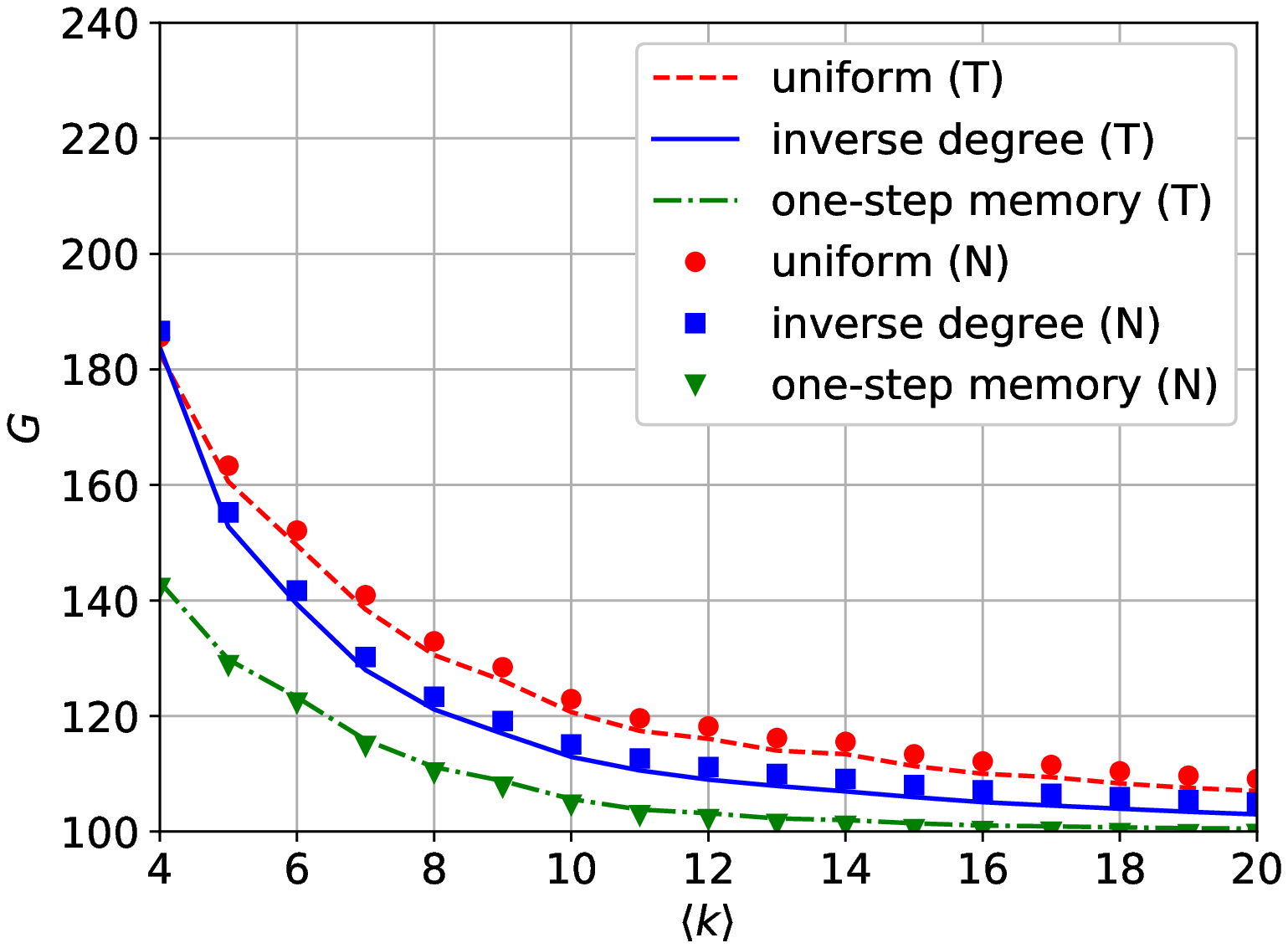}
    }\\ 
    \subfloat[WS networks \label{fig:GMFPTWSNet}]{
		\includegraphics[width=0.9\columnwidth]{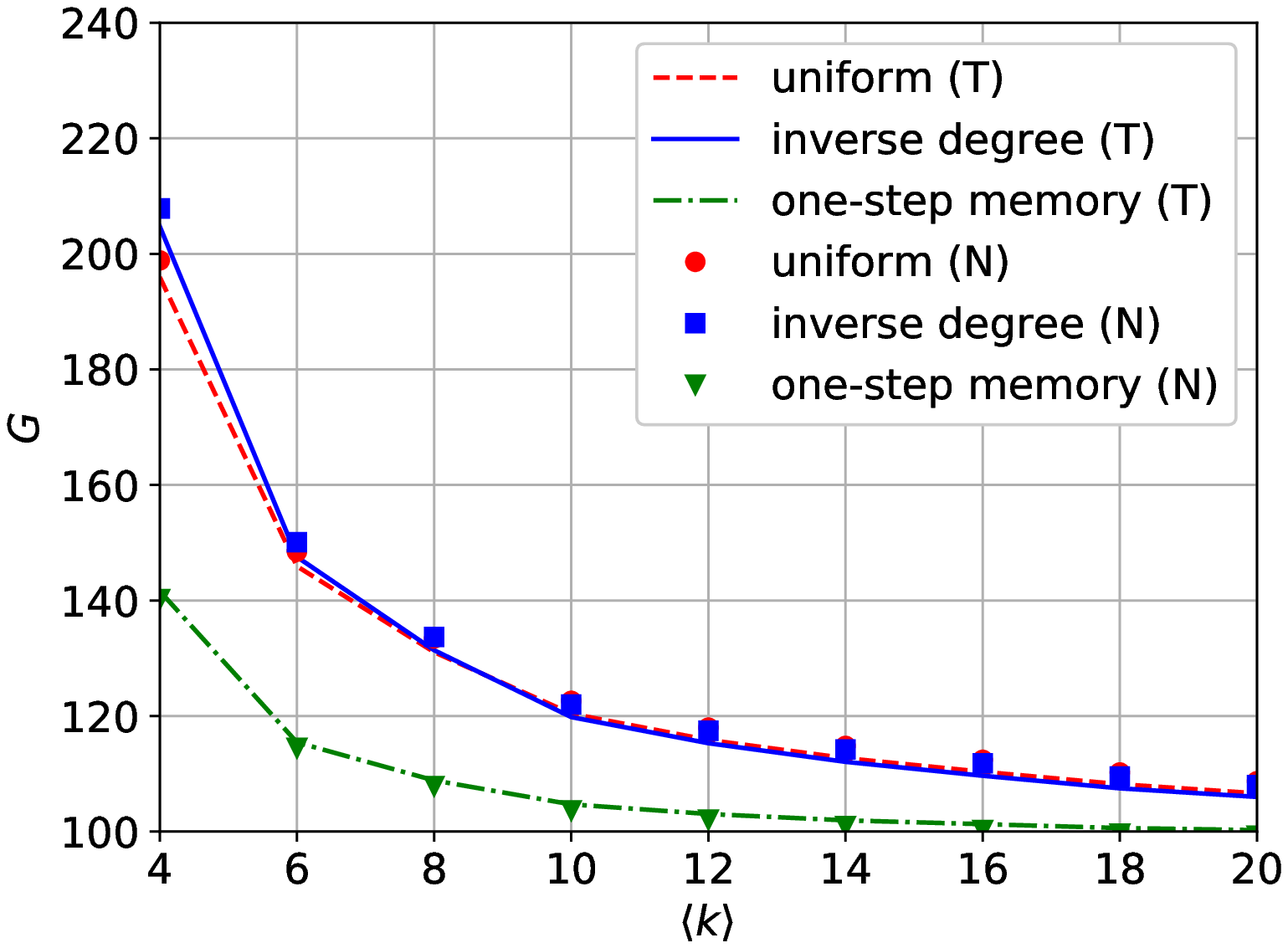}
	}
    \caption{GrMFPT in (a) BA, (b) ER, and (c) WS networks of $N=100$ nodes with different average node degree $\langle k \rangle$ for the three cases: uniform (red line/circle), inverse degree (blue line/square) and one-step memory (green line/triangle). The lines are theoretical values (T) and the markers numerical estimates (N).}
    \label{fig:GMFPTNet}
\end{figure}

In Figure \ref{fig:GMFPTBANet} we compare the obtained GrMFPTs for the URW, inverse-degree-biased random walk and the memory-based one over scale-free networks. The horizontal axis represents the average degree $\langle k \rangle$ which is approximately $2m$, where $m \in [2,10] $. The seed network is composed of $m_0=m$ nodes without edges. First, one can observe that the numerical (N) and the theoretical (T) results are very close, which confirms the correctness of the analytical expressions. The memory-based random walk always outperforms the uniform one. The inverse-degree-biased random walk is also better than the URW, when the average node degree is not very small. One can notice that all curves decrease asymptotically towards the value corresponding to the number of nodes $N$. As we will see from the other numerical results, $N$ seems to be the minimal possible value for the GrMFPT. Thus, as optimal random search could be considered the one for which GrMFPT equals the number of nodes, $G=N$. Although for networks with very large average degree the GrMFPT seems to approach to $N$ for different kinds of random walk, the effectiveness of a biasing procedure becomes apparent for less connected networks. 

We note that there is deterministic strategy that is twice faster and which holds for graphs that have a Hamiltonian cycle. It is a walk passing though all nodes and visiting each node only once. We emphasize here that determination whether a graph has a Hamiltonian cycle is not a trivial task and was proven to be an NP-complete problem \cite{karp1972reducibility}. In that case the MFPT from the source to the target will equal the number of nodes in between them along the cycle, and for uniformly chosen starting and target node, one can easily show that GrMFPT will be $N / 2$.

The two biasing procedures bring search improvement for the purely random Erd\H{o}s-R\'enyi graphs also, as it is shown in Figure \ref{fig:GMFPTERNet}. We generated $10$ network instances with $N=100$ nodes for different average node degree $\langle k \rangle$ by varying the link existence probability $p \in [0.04, 0.2]$. 
As it can be seen the inverse-degree biasing gives lower GrMFPT than the URW, except for $\langle k \rangle = 4$ where they are about the same, while the one-step memory outperforms them both. Again the numerical results are in accordance with the theoretical ones.

In Figure \ref{fig:GMFPTWSNet} we show how the biasing affects the random walk in Watts-Strogatz networks, where the rewiring probability is $p=0.2$. Unlike for the other network types under study, the inverse-degree biasing does not improve the GrMFPT. This is probably due to the smaller degree variability in this kind of networks. On the other hand, the one-step memory approach still reduces the GrMFPT, as it was the case for the other network types. The theoretical expressions are once again confirmed by the numerical simulations.

\begin{figure}[!tb]
	\subfloat[BA networks \label{fig:KL_BA}]{
		\includegraphics[width=0.9\columnwidth]{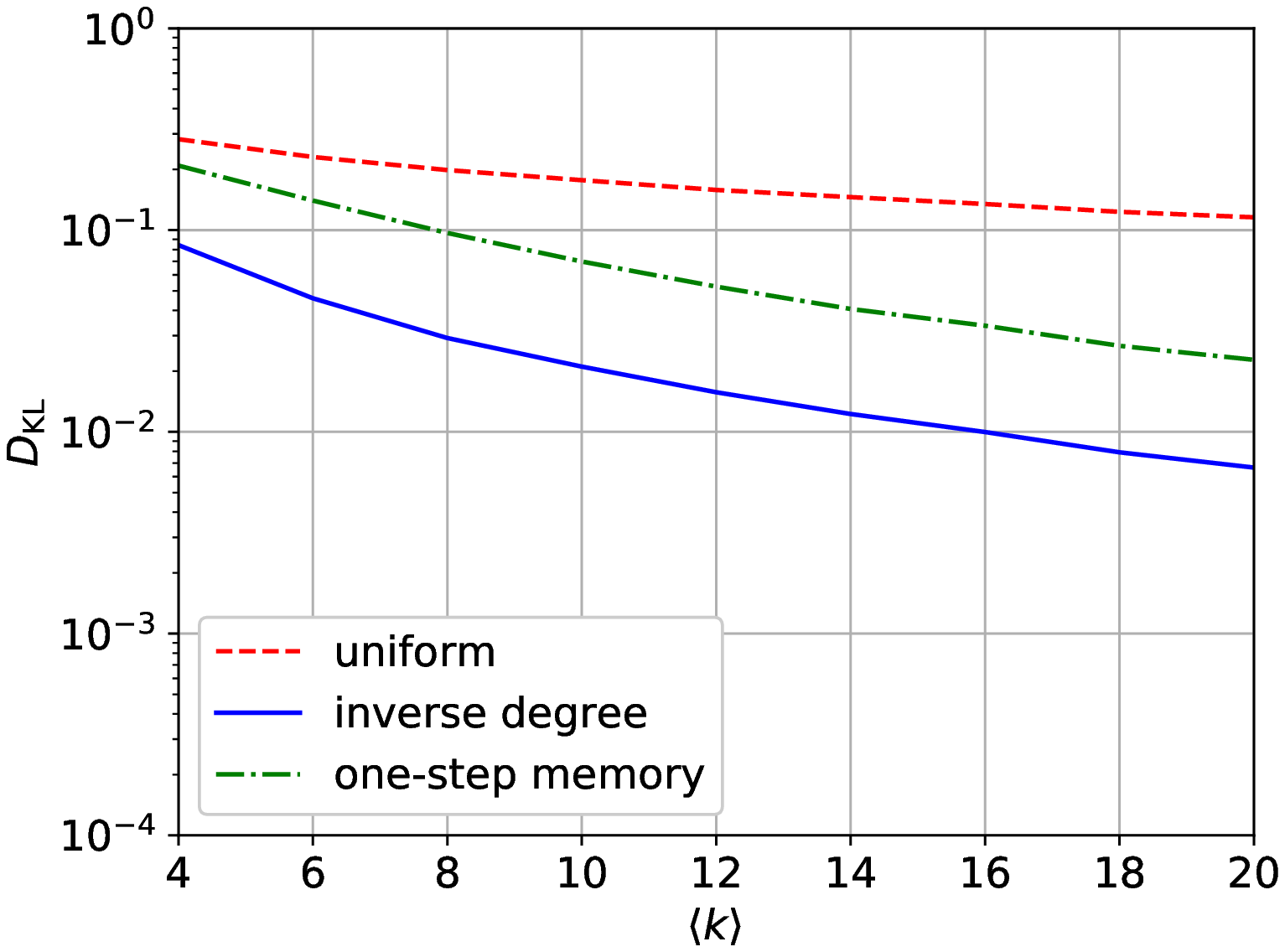}
	}\\
	\subfloat[ER networks \label{fig:KL_ER}]{
		\includegraphics[width=0.9\columnwidth]{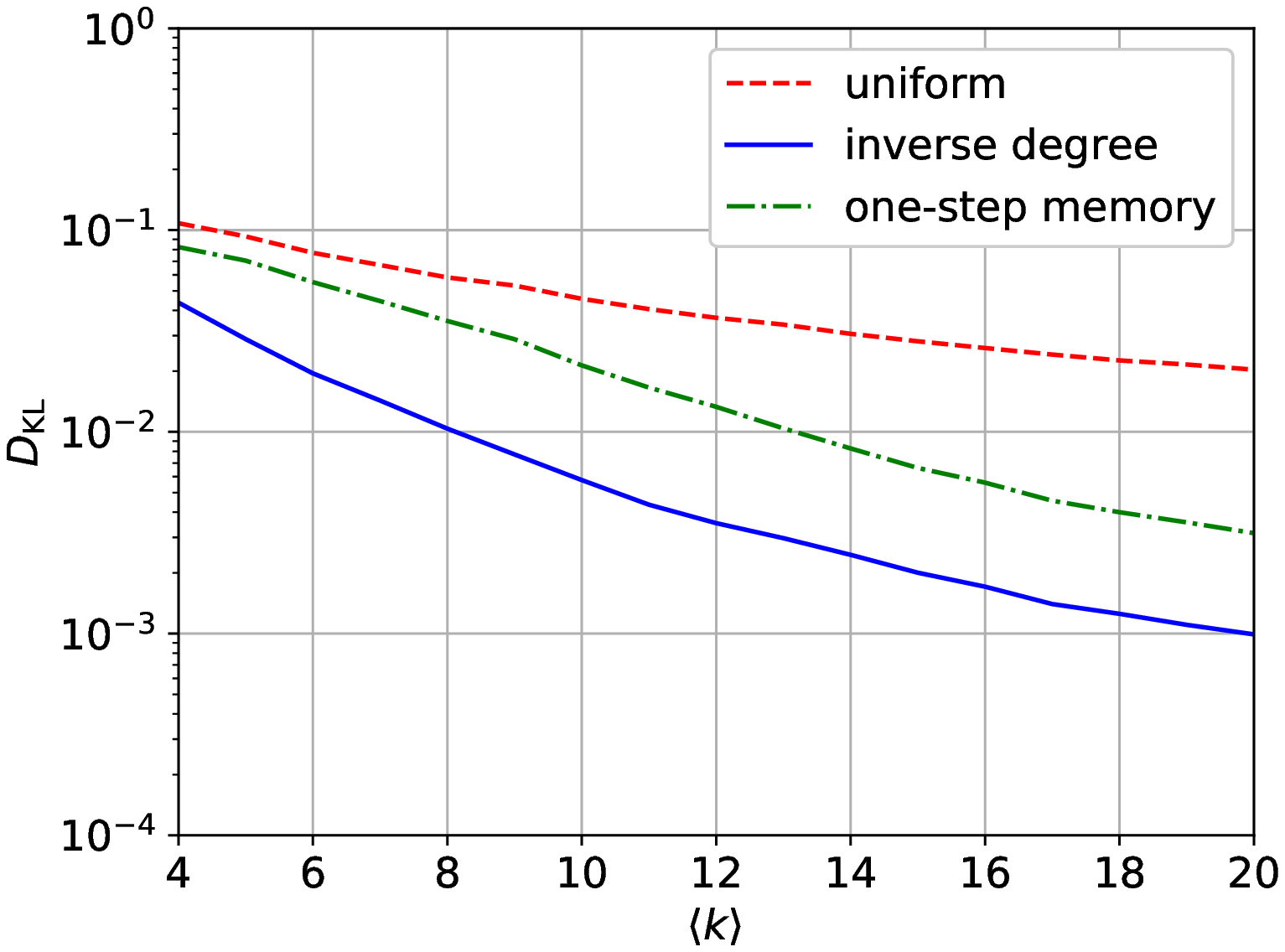}
	}\\
	\subfloat[WS networks \label{fig:KL_WS}]{
		\includegraphics[width=0.9\columnwidth]{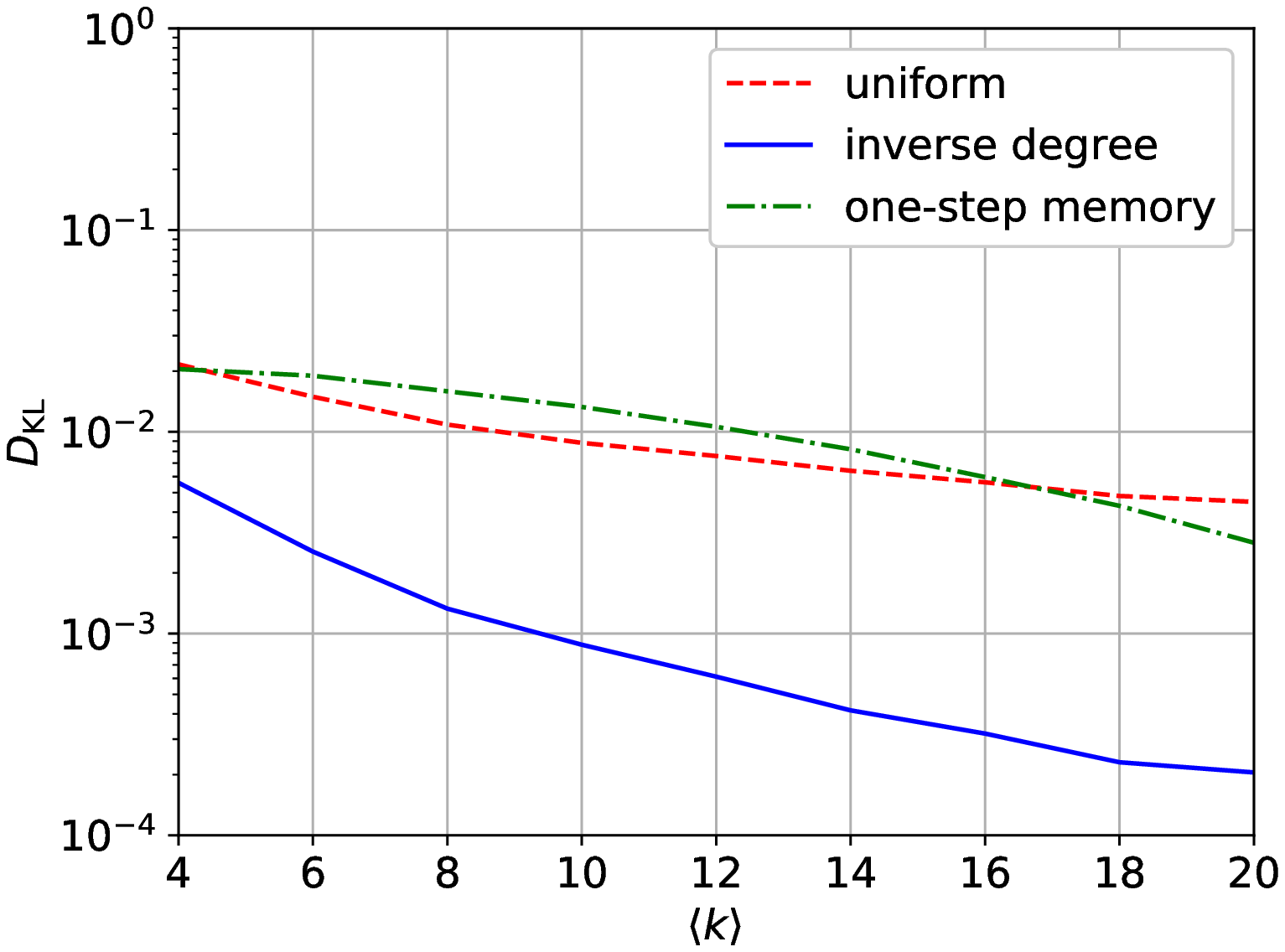}
	}
\caption{Kullback-Leibler divergence of the stationary occupation probability of uniform (red), inverse degree (blue), and one-step memory (green) random walks from the uniform density in (a) BA,  (b) ER, and (c) WS networks with $N=100$ nodes for different average node degrees.
\label{fig:KL_BA_ER}}
\end{figure}

We also made numerical experiments to see whether a mechanism behind the search improvement is nearly uniform stationary occupation probability. The extent of flattening of the stationary occupation probability was studied with the Kullback-Leibler (KL) divergence \cite{mackay2003information}. KL divergence estimates the deviation of one distribution from another. In the case when one has two discrete distributions $P(i)$ and $Q(i)$, it is defined as
\begin{equation}
    D_{\textrm{KL}}(P||Q) = \sum_{i} P(i) \log \frac{P(i)}{Q(i)}.
\end{equation}
One can notice from the definition that this is asymmetric quantity, $D_{\textrm{KL}}(P||Q) \neq D_{\textrm{KL}}(Q||P)$, and within the definition provided above, $P$ has the role of the prior, or the distribution with which we compare. In our case it is the constant $P(i) = 1/N$. This divergence vanishes when the two distributions coincide. In Figure \ref{fig:KL_BA} is shown the KL divergence between the constant density and those for the uniform, inverse degree and one-step memory random walks in BA networks. As can be noticed, both biasing procedures result in invariant density that is closer to the flat one, than the uniform approach does. Also, the larger the average degree is, the approximation of the invariant density with the uniform one is more correct, as the theoretical analysis in the Appendix suggests. However, even though for networks with smaller average degree the biasing makes the distribution closer to the uniform, searching is slower than for the URW. This clearly indicates that the leveling of the visiting frequencies is not always sufficient for optimizing the search.  

Similarly, Figure \ref{fig:KL_ER} shows the KL divergence between the uniform density and the distribution of the visiting frequency for the URW and the two other random walks in ER networks. Once again, the biasing yields a density that is closer to the constant one than the URW, which is probably the reason for the lower GrMFPT obtained in Figure \ref{fig:GMFPTERNet}. On the other hand, in WS networks (see Figure \ref{fig:KL_WS}) the inverse-degree biasing gives a density which is closer to a constant than the URW, while the one-step memory approach does not, even though it proved fastest in such scenario as is evidenced in Figure \ref{fig:GMFPTWSNet}.

\begin{figure}[!tb]
	\subfloat[GrMFTP \label{fig:GMFPTERDirect}]{
		\includegraphics[width=0.9\columnwidth]{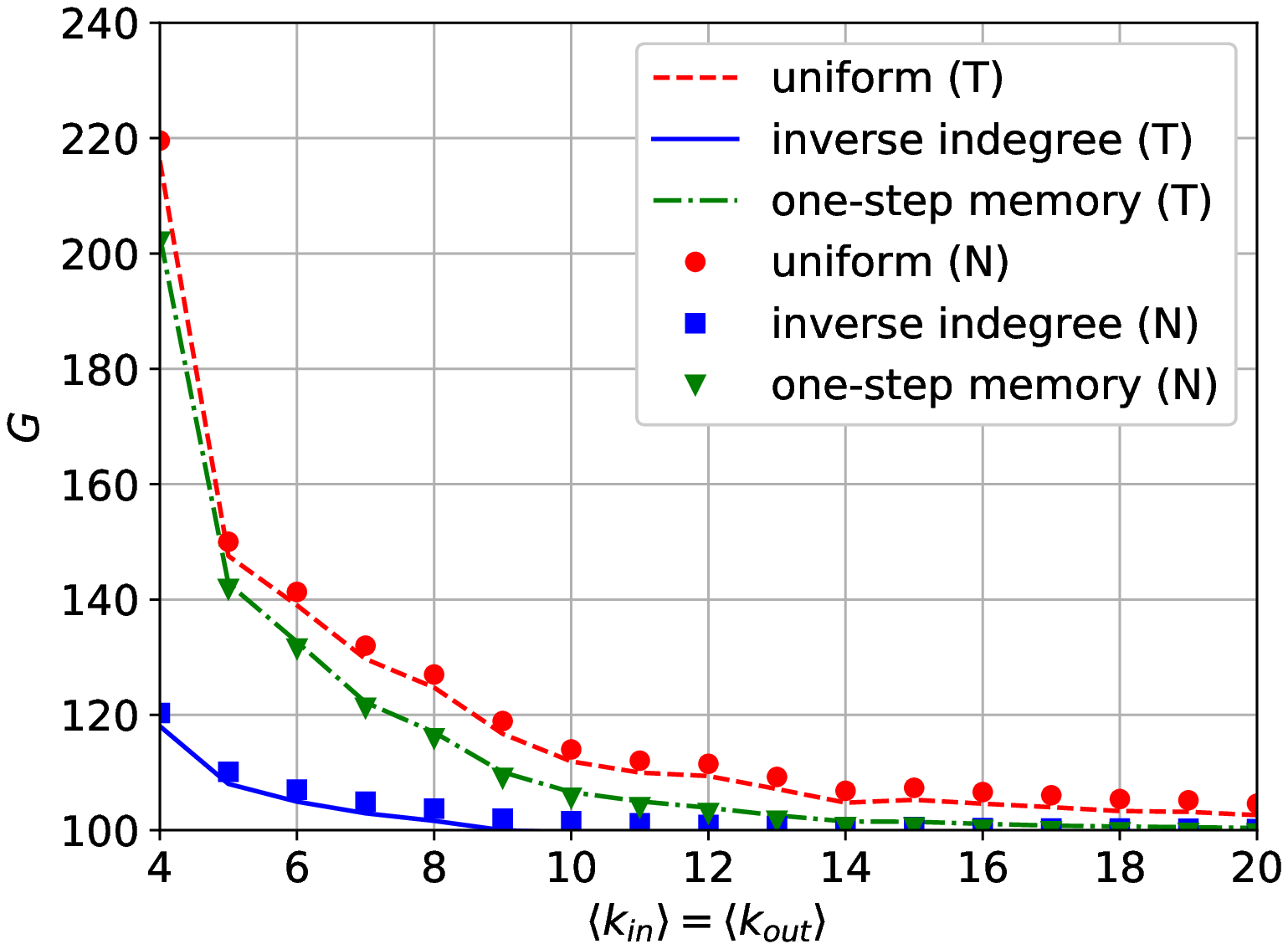}
	}\\
	\subfloat[Kullback-Leibler divergence \label{fig:KL_ER_directed}]{
		\includegraphics[width=0.9\columnwidth]{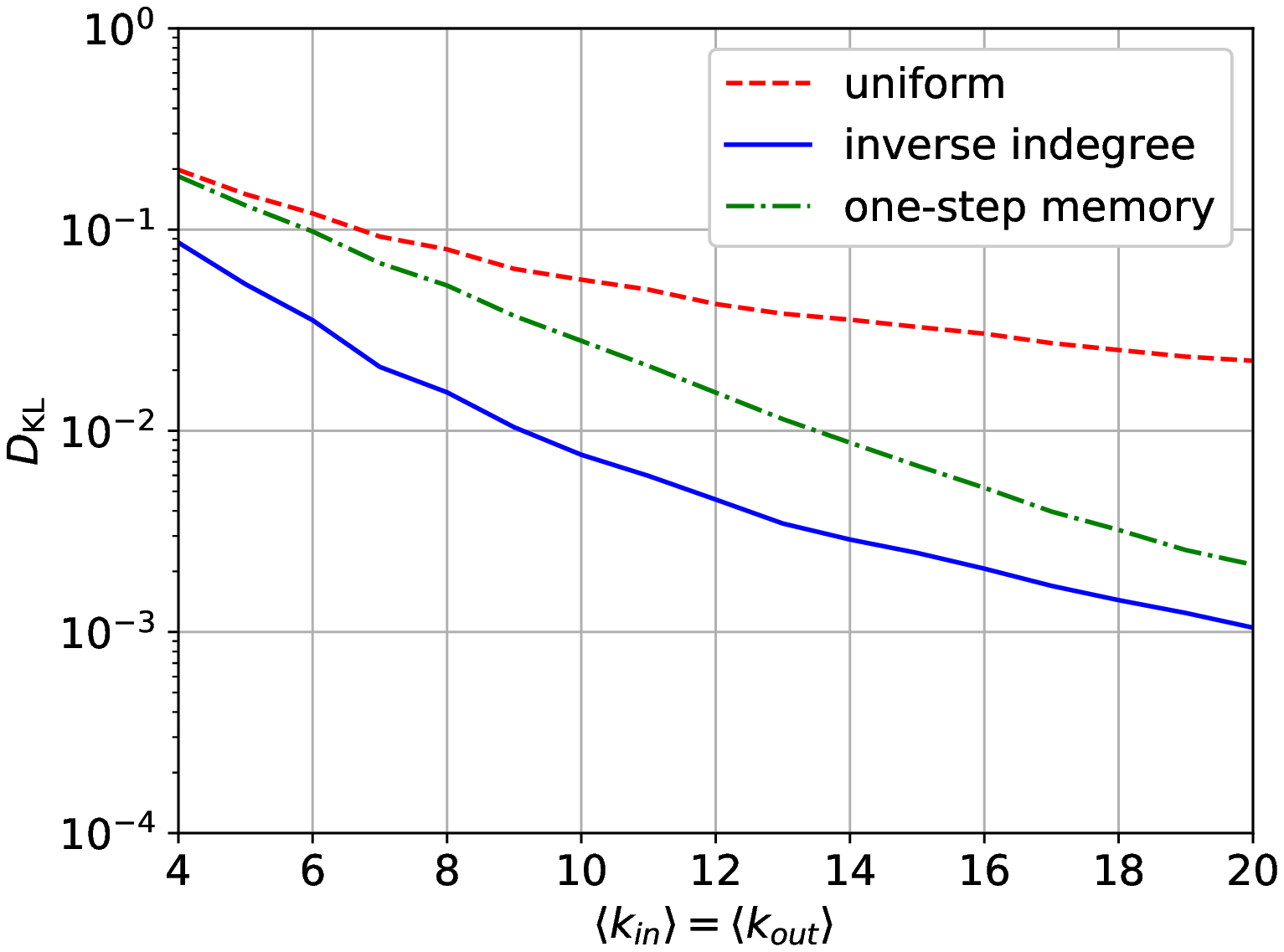}
	}
	\caption{Random walks in directed ER networks with different average degree $\langle k \rangle$: (a) Comparison of the GrMFPT for uniform (red circles), inverse indegree (blue squares) and one-stop memory (green triangles); and (b) Kullback-Leibler divergence of the invariant density from a uniform density for three approaches: uniform (red circles), inverse-indegree-biased (blue squares) and one-step memory (green triangles).
	\label{fig:ERDirect}}
\end{figure}

We also numerically compared the searching performance of the three kinds of random walk in directed networks. In the Figure \ref{fig:GMFPTERDirect} are shown the respective GrMFPTs. We can see that the one-step memory provides better results than the URW, but the inverse indegree approach outperforms them both significantly, which was not the case in the undirected networks. Biasing based on inverse outdegree performs slower than the URW (results are not shown), as it is expected.

The flattening of the invariant density is an ingredient which helps in search improvement in directed networks as well. We have numerically verified that, as expected, for well connected networks when biasing of random walk is based on inverse of indegrees, the invariant density is closer to the constant, than that of a URW. In Figure \ref{fig:KL_ER_directed} are shown the KL divergence of the URW on directed ER networks with the two biasing alternatives: one based on inverse of indegrees, and another on the one-step memory. The results are in concordance with the theoretical analysis.

We have finally tried the searching performance of the three approaches in two real world networks. The first network is a snapshot of the Internet topology at autonomous systems level obtained from BGP logs on 2.1.2000, which is an undirected graph consisting of $6474$ nodes and $13233$ links \cite{leskovec2005graphs}. Its average node degree is $\langle k \rangle \approx 4$. The second network is an extracted set of Wikipedia pages \cite{west2009wikispeedia,west2012human}. The graph is directed and consists of $4592$ nodes and $119882$ links, from which we take the largest strongly connected component that has $4051$ nodes and $119000$ links. The average indegree and outdegree of the largest component are $\langle k_\mathrm{in} \rangle = \langle k_\mathrm{out} \rangle \approx 29$. These networks are larger and it is much more difficult to calculate the GrMFTP theoretically, so in Table \ref{tab:1} we provide only the results obtained by numerical simulations. The results for the Internet network are obtained by averaging over $10^6$ randomly selected source-target pairs out of $6474 \times 6473$ possible couples. For the extract of the Wikipedia network the averaging is performed with $1.5 \cdot 10^5$ pairs, out of $4051 \times 4050$ possible, as the simulations take much longer due to the larger number of steps required to reach the targets. One can note that for the undirected case the inverse-degree biasing worsens the search of the URW, because majority of nodes are not well connected as the theory asks, while it shows great reduction of the MFPT for the directed network. The memory-based strategy performs well in both scenarios. These results confirm our previous findings for paradigmatic network models that the inverse indegree biasing is better for directed networks, while the memory-based approach outperforms the others for undirected ones.

\begin{table}[h]
		\caption{GrMFPT for two real networks with uniform, inverse degree and one step-memory random walks.\newline}
		\label{tab:1}
		\begin{tabular}{l|l|l|l}
			\hline
			Network 			& Uniform 	& Inverse degree 	& One-step memory\\
			\hline
    		Internet(AS)		& $1.93 \cdot 10^4$     & $1.78 \cdot 10^5$	& $1.80 \cdot 10^4$\\
			Wikipedia(extr.)  	& $3.01 \cdot 10^7$ 	& $1.09 \cdot 10^4$	& $8.15 \cdot 10^5$\\
		\end{tabular}
\end{table}

\section{Conclusions}\label{SEC:conclusions}

In this work we studied random walk on complex networks with transition probabilities that depend on the nodes visited in the recent past. We have shown that such walks can be analyzed with appropriate Markov chain, and for the case of memory of one step it was derived an exact expression for the MFPT between pairs of nodes. One particular navigation algorithm was proposed that avoids the hubs by accounting for the two-hop-paths between the nodes. The searching ability of this algorithm was compared to that of the URW, and of another hubs-avoiding biased random walk with jumping probabilities inversely proportional to the node degrees. The proposed one-step memory approach has shown better searching performance than the URW and the inverse-degree-biased random walk for undirected networks, particularly when the majority of nodes have a small degree. We have furthermore demonstrated that the inverse-degree biasing based on indegree, leads to improved random search in directed networks, which is even better than the memory-based one. The introduced technique with absorbing Markov chain could be also applied in theoretical analysis of other scenarios. One example is random searching of target when each node knows the identity of its neighbors. In this case the absorbing states would be all neighbors of the target.

The numerical experiments on generic network models besides verifying the correctness of the theoretical expressions, have shown that when the nodes have sufficiently enough neighbors, the GrMFPT approaches the number of nodes from above, for the three considered kinds of random walk. However, the usefulness of the biasing alternatives is that they allow achieving nearly optimal performance for less connected networks than the URW does. Also, both biasing approaches show better flattening towards the constant of the stationary occupation probability than the URW. The inverse-degree biasing results in stationary occupation probability that is always closer to the uniform than the two other kinds of random walk. This is not sufficient for best searching because it was obtained that the memory-based random walk performs better on undirected networks. However, the obtained results suggest that leveling of the stationary occupation probability can at least serve as an indicator for possibly good searching algorithm, particularly when the respective KL divergence has very small value.



\section*{Acknowledgement}
This research was partially supported by the Faculty of Computer Science and Engineering, at the SS Cyril and Methodius University in Skopje, Macedonia. 

\appendix* 

\section{Conditions for nearly uniform distribution of the visiting frequency}
\label{appendix}

The analysis in this section will be performed for random walk on directed complex networks, although the same reasoning applies for undirected networks as well with minor modifications. Consider random walk on directed network, with the transition probability toward certain node $j$ to be inversely proportional to its indegree $k_j^{\mathrm{in}}$. Due to the normalization, the jumping probability from node $i$ to its neighbor $j$ would then be
\begin{equation}
p_{ij} = \frac{1 / k_j^{\mathrm{in}}} {\sum_{l \in \mathcal{N}_i^{\mathrm{out}}} 1 / k_l^{\mathrm{in}}},
\label{eq:P_inv_indegree}
\end{equation}
where $\mathcal{N}_i^{\mathrm{out}}$ denotes the set of neighbors of the node $i$ toward which it points to.
Define node-centric, local average of the reciprocal of indegrees of the neighbors as
\begin{equation}
    \left<1 / k \right>_i^{\mathrm{in}} = \frac{1}{k_i^{\mathrm{out}}} \sum_{l \in \mathcal{N}_i^{\mathrm{out}}}1 / k_l^{\mathrm{in}}, \label{eq:local_in_deg_avg}
\end{equation}
where the subscript $i$ in the average denotes that it is calculated only over the set $\mathcal{N}_i^{\mathrm{out}}$. Then the normalization sum in Eq. (\ref{eq:P_inv_indegree}) can be expressed through the local average as
\begin{equation}
\sum_{l \in \mathcal{N}_i^{\mathrm{out}}} 1 / k_l^{\mathrm{in}} = k_i^{\mathrm{out}} \left<1 / k \right>_i^{\mathrm{in}}.
\end{equation}
Now, consider well connected uncorrelated networks. Such networks are those where the degree of any node is independent on the degrees of its neighbors and where for majority of the nodes holds $k_i^{\mathrm{in}} \gg 1$ and $k_i^{\mathrm{out}} \gg 1$. Then the local average can be approximated with the network average of the reciprocal of indegrees
\begin{equation}
    \left<1 / k \right>_i^{\mathrm{in}} \approx \left<1 / k \right>^{\mathrm{in}} = \frac{1}{N} \sum_{j=1}^N 1/k_j^{\mathrm{in}}
    \label{eq:avg_in_degree_approx}
\end{equation}
In such case the normalization sum appearing in the denominator in (\ref{eq:P_inv_indegree}) can be conveniently expressed through the network average as
\begin{equation}
\sum_{l \in \mathcal{N}_i^{\mathrm{out}}}1 / k_l^{\mathrm{in}} \approx k_i^{\mathrm{out}} \left<1 / k \right>^{\mathrm{in}}.
\label{eq:Inv_in_deg_approx}
\end{equation}
The stationary distribution of the visiting frequency satisfies the following set of self-consistent equations
\begin{equation}
    w_j = \sum_{i \in \mathcal{N}_j} p_{i,j} w_i,
\end{equation}
for each node $j$. This means that the following holds
\begin{equation}
w_j = \sum_{i \in \mathcal{N}_j^{\mathrm{in}}} \frac{1 / k_j^{\mathrm{in}}}{k_i^{\mathrm{out}}\left<1/k\right>^{\mathrm{in}}} w_i = \frac{1 / k_j^{\mathrm{in}}}{\left<1/k\right>^{\mathrm{in}}}  \sum_{i \in \mathcal{N}_j^{\mathrm{in}}} \frac{w_i}{k_i^{\mathrm{out}}}. \label{eq:Stat_dens_indegree}
\end{equation}
If one assumes that the invariant density is constant $w_i = 1/N$, then from Eq. (\ref{eq:Stat_dens_indegree}) one would have
\begin{equation}
\frac{1}{N} \approx \frac{1 / k_j^{\mathrm{in}}}{N\left<1/k\right>^{\mathrm{in}}}  \sum_{i \in \mathcal{N}_j^{\mathrm{in}}} \frac{1}{k_i^{\mathrm{out}}}. \label{eq:Stat_dens_approx_indeg}
\end{equation}
Now, for networks where the direction of the links is independent on the degree of nodes, the averages of reciprocals of indegrees and outdegrees would be nearly the same
\begin{equation}
  \left<1 / k \right>^{\mathrm{in}} \approx \left<1 / k \right>^{\mathrm{out}}.
\end{equation}
For networks where most of the nodes have many incoming and outgoing links, one can make the following approximation
\begin{equation}
    \sum_{i \in \mathcal{N}_j^{\mathrm{in}}} \frac{1}{k_i^{\mathrm{out}}} \approx k_j^{\mathrm{in}} \left<1 / k \right>^{\mathrm{out}} \approx k_j^{\mathrm{in}} \left<1 / k \right>^{\mathrm{in}}.
    \label{eq:sum_inv_ingedree}
\end{equation}
Plugging the last approximation in the stationary density equation (\ref{eq:Stat_dens_approx_indeg}), one will see that it is identity. 

We should mention that although network averages of the reciprocals of in- and outdegrees are nearly equal, the biasing inverse to the outdegrees does not result in a stationary distribution approaching to uniform one. The reason for that is the fact that the sum of inverse of degrees (\ref{eq:sum_inv_ingedree}) is always proportional to the indegree of the node $j$ because it accounts for neighbors pointing to the node $j$. By repeating the analysis above by using biasing with inverse of outdegrees, one can verify that the stationary density condition like (\ref{eq:Stat_dens_approx_indeg}) is not satisfied. 
%

\end{document}